\documentclass[]{aa}
\usepackage{graphicx}
\usepackage{amsmath}
\usepackage{txfonts}
\usepackage{natbib}
\begin{document}
   \title{An Effective temperature calibration for solar type stars using equivalent width ratios}

   \subtitle{A fast and easy spectroscopic temperature estimation}

   \author{S. G. Sousa\inst{1,}\inst{2}
          \and
          A. Alapini\inst{2,}\inst{3}
          \and
          G. Israelian\inst{2,}\inst{4}
          \and
          N. C. Santos\inst{1}
	    }
   \offprints{S. G. Sousa: sousasag@astro.up.pt}

	  \institute{Centro de Astrof\'isica, Universidade do Porto, Rua das Estrelas, 4150-762 Porto, Portugal
	  \and	Instituto de Astrof\'isica de Canarias, 38200 La Laguna, Tenerife, Spain
	  \and School of Physics, University of Exeter, Stocker Road, Exeter EX4 4QL, United Kingdom
	  \and Departamento de Astrofisica, Universidade de La Laguna, E-38205 La Laguna, Tenerife, Spain
     }
  
   \date{}

 
  \abstract
   {}
   {The precise determination of the stellar effective temperature of solar type stars is of extreme importance for Astrophysics. We present an effective temperature calibration for FGK dwarf stars using line equivalent width ratios of spectral absorption lines.}
   {The ratios of spectral line equivalent width can be very sensitive to effective temperature variations for a well chosen combination of lines. We use the automatic code ARES to measure the equivalent width of several spectral lines, and use these to calibrate with the precise effective temperature derived from spectroscopy presented in a previous work.}
   {We present the effective temperature calibration for 433 line equivalent width ratios built from 171 spectral lines of different chemical elements. We also make available a free code that uses this calibration and that can be used as an extension to ARES for the fast and automatic estimation of spectroscopic effective temperature of solar type stars.}
   {}


   \maketitle
%

\section{Introduction}

One of the most important parameters in stellar Astrophysics is the effective temperature. However this parameter is very difficult to be measure with high accuracy, specially for stars that are not so closely related to our very own Sun. Also the correct (or incorrect) determination of this parameter will have a major effect on the determination of other associated parameters such as the surface gravity and the chemical composition of the associated star.

There are several methods that one can choose in order to derive precisely the effective temperature of stars. There is the Photometry way that is usually based on the calibration of several band colors, such as (B-V), (b-y), (V-K), etc... \citep[e.g.][]{Nordstrom-2004}. Spectroscopy is another very powerful technique to determine the effective temperature. In this case (e.g. studying the H$\alpha$ wings or using the iron lines to find excitation and ionization equilibrium) a very careful analysis of the stellar spectra is needed together with the comparison with stellar atmosphere models. Spectroscopy is a very precise method, although it has the disadvantage to be a very time-consuming method when using the standard procedure. Interferometry is an almost direct method to derive the temperature that relies on the accurate determination of the stellar angular diameters. These are ultimately combined with the bolometric flux of the stars. Although there are still the use of models in this method (e.g. limb-darkening and prior knowledge of the calibration stars), this method can help to find the answer for the problem of the temperatures scales for the other different methods. However, the measurement of interferometric radius is only possible for a small number of stars that are very close and/or very bright.


In this work we will present a different approach using spectroscopy, where we can fastly estimate the temperature with high precision through line ratios. However, we use the equivalent widths instead of the line depths that were used before in similar line ratios works. The equivalent width has intrinsically more information then line depths and can show significant differences in the calibrations. Nevertheless our work is inspired, as similar works, in the following statement: "There is no doubt that spectral lines change their strength with temperature, and the use of the ratio of the central depths of two spectral lines near each other in wavelength has proved to be a near optimum thermometer" \citep[][]{Gray-2004}. Therefore the ratio of the equivalent of two lines that have different sensitivity to temperature is an excellent diagnostics for measuring the temperature in stars or to check small temperature variations of a given star. Although the effective temperature scale can still be questioned, as it happens with other methods, this method allows us to reach a precision for the temperature down to a few Kelvin in the most favorable cases \citep[e.g.][]{Gray-1991,Strassmeier-2000,Gray-2001,Kov-2003}. In this work we will use the equivalent width (EW) to measure the line strength. Normally the line depth is used instead, therefore we make a small discussion, further ahead in the article, about the reasons for our choice in EWs.
 
In this work we will calibrate this line-ratio technique using data from our previous work \citep[][]{Sousa-2008} where we have determined precise spectroscopic stellar parameters for a large sample of solar-type stars using high resolution spectra observed with HARPS. The calibration is used in a code that can be used as an extension to ARES to quickly estimate spectroscopic effective temperatures for FGK dwarf stars. In chapter 2 we describe the data used in this work. In Chapter 3 we describe the procedure to determine the calibration and explain how the lines and the line-ratios where chosen and calibrated. In Chapter 4 we present a possible procedure to estimate the effective temperature using the derived calibration. We also present a simple code that uses this calibration and this process to derive the effective temperatures. We then test this code and calibration in Chapter 5. We finalize this Chapter testing the uncertainties of this procedure when taking into account the errors coming from the measurements of the continuum level of the spectra, and assuming different spectral wavelength intervals. In Chapter 6 we summarize and conclude.



\section{The Data}

\begin{figure}[!ht]
\centering
\includegraphics[width=8cm]{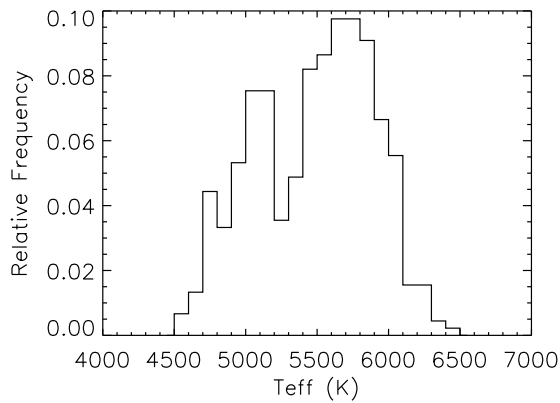}
\caption[]{Teff distribution of the calibration sample.}
\label{fig1}
\end{figure}

In this work we use the same data as presented in the work of \citet[][]{Sousa-2008}. The sample of stars is composed of 451 solar-type stars with high resolution spectra (R $\sim$ 110000) observed with HARPS. The S/N from this sample varies from 70 to 2000, with 90\% of the spectra having a S/N higher than 200. Precise spectroscopic stellar parameters (Teff, log g, [Fe/H]) were obtained in \citet[][]{Sousa-2008} for all the stars in the sample. Figure \ref{fig1} shows the effective temperature distribution of our sample. This is an ideal data set to fulfill a line-ratio calibration, specially because these stars were observed and analyzed in a consistent and systematic manner. Moreover it was presented a series of comparisons with results coming from other methods for the determination of the temperature and the conclusion is that they show a good consistency. For further details about the sample of stars and respective stellar parameters determination see \citet[][]{Sousa-2008}.

\section{Teff vs. EW line-ratios - An empirical calibration}

In this section we describe how the line-ratios were selected and how the calibration was derived. We start with a large list of lines for several chemical elements, then we select a subset of line-ratios from all the possible combinations that can be combined with this list and we finally present the selection of the best line-ratios with an empirical effective temperature calibration.

\subsection{Line List}

In order to choose the list of the lines that we will use to analyze the spectra, we started with the large line list of iron lines used in the previous work in \citet[][]{Sousa-2008} for a detailed spectroscopic analysis. Moreover, to include lines of different elements we added the list of lines presented in \citet[][]{Neves-2009} -- in which work the chemical abundance of several species was derived for the same 451 stars. Finally we also added 19 lines that were not included in the last list and that are usually used for chemical abundance determination. This additional lines are presented in Table \ref{tab1}. Therefore the preliminary large line list used to build our first set of line-ratios is composed of a total of 498 spectral lines.

\begin{table}[!t]
\centering 
\caption[]{Spectral lines added to the lists of lines presented in \citet[][]{Sousa-2008} and in \citet[][]{Neves-2009}.}
\begin{tabular}{lccc}
\hline
\hline
\noalign{\smallskip}
$\lambda$ (\AA) & $\chi_{l}$ & $\log{gf}$ & Ele.\\
\hline

6318.72&	5.11&	-1.996&	MgI\\
5621.60&	5.08&	-2.500&	SiI\\
5690.43&	4.93&	-1.790&	SiI\\
6414.99&	5.87&	-1.100&	SiI\\
6583.71&	5.95&	-1.640&	SiI\\
6604.60&	1.36&	-1.160&	ScII\\
5866,45&	1.07&	-0.840&	TiI\\
6330.13&	0.94&	-2,92&	CrI\\
5703.59&	1.05&	-0.460&	VI\\
5727.05&	1.08&	0.000&	VI\\
6039.73&	1.06&	-0.650&	VI\\
6090.21&	1.08&	-0.150&	VI\\
6111.65&	1.04&	-0.715&	VI\\
6135.36&	1.05&	-0.746&	VI\\
6216.35&	0.28&	-0.900&	VI\\
5578.72&	1.68&	-2.650&	NiI\\
5682.20&	4.10&	-0.39&	NiI\\
5754.68&	1.94&	-2.33&	NiI\\
6007.31&	1.68&	-3.330&	NiI\\

\hline
\end{tabular}
\label{tab1}
\end{table}

\subsection{First selection of the line-ratios}

Since we are looking for an empirical relation of each line-ratio as a function of temperature it is only natural that we must choose the appropriate combination of lines in order to be more sensitive to the variation of temperature and to be as much as possible independent from other factors. This factors can be either physical ones -- such as due to the metallicity abundances or the surface gravity differences, or non physical ones -- such as the one coming from the subjective measurements of the equivalent widths.

In order to avoid possible systematics in the line-ratio calibrations we present here some ground rules for the preliminary selection of the line-ratios. Some of these rules were also presented in other similar works such as the one from \citet[][]{Kov-2003}. 
\begin{itemize}
\item The first condition is to only use line-ratios composed with lines that are close together in the wavelength domain. Therefore we will restrict our line-ratios to be built with lines that are less than 70\AA\ away from each other. This value is the one used in \citet[][]{Kov-2003}. This condition aims at eliminating possible errors coming from the continuum determination in the measurement of the equivalent widths for these lines.

\item The second condition is to only allow lines that have an excitation potential difference greater than at least 3 eV. In this way we are compiling line-ratios that will be more sensible to effective temperature variations. This is true because the equivalent width of the lines with higher excitation potential will change faster with temperature than the ones from lower excitation potential lines \citep[][]{Gray-1994}.


\end{itemize}
Using the above two conditions we end up with a total of 934 line-ratios that should be sensible to temperature. Any other physical dependence such as the ones we can expect due to surface gravity and metallicity differences will be seen in the dispersion of each line-ratio calibration. The line-ratios that will present a larger dispersion, when compared to the average dispersion from all line-ratios, will not be used for the final list of calibrated line-ratios.

\subsection{Automatic measurement of the EWs with ARES}

In this work we will use the equivalent width (EW) of a line to measure their strength. Measuring the equivalent width is very subjective because the standard process is based in interactive routines where the continuum position is normally fitted by eye for each individual line. Therefore, each measurement has an intrinsic error that is very difficult to estimate and is often ignored. Using an automatic process to obtain these EW measurements will eliminate a large part of these subjective errors, since it will be possible to measure the lines in a consistent and systematic way.

ARES \citep[][]{Sousa-2007} is a code that was developed to automatically perform this task. This automatic code measures EWs in absorption spectra and it has recently been proven \citep[][]{Sousa-2008} to be a very powerful tool when there is a large quantity of data to analyze. \footnote{http://www.astro.up.pt/$\sim$sousasag/ares}

The equivalent widths in this work were computed following the same procedure as in \citet[][]{Sousa-2008}. The most important parameter from ARES is the ''rejt'' input parameter that depends strongly on the spectrum S/N. The value in this parameter is used by ARES to select the points for the determination of the local continuum setting the level of its position (See \citet[][]{Sousa-2007} for more details). We follow Table 2 of the work of \citet[][]{Sousa-2008} to choose the appropriate value for the ''rejt'' parameter accordingly to the S/N of the spectra.

\subsection{Calibration for each line-ratio}

In order to find the calibration for each line-ratio we used the effective temperatures presented in \citet[][]{Sousa-2008} and plotted them against each line-ratio value for all the 451 calibration stars. In order to fit this empirical relation we chose a 3rd order polynomial function allowing us to model Teff vs. EW ratio with a minimal number of free parameters (4).

The first condition for each line-ratio calibration is to only use the EW line-ratio values greater than 0.01 and lower than 100. This is to avoid ratios composed of a very weak line and/or a very strong line. For these cases we will have a higher uncertainty on the EW measurement because on one hand the continuum will have a larger influence on the smaller lines and, on the other hand, ARES Gaussian fit to the spectral line will not be the most appropriate fitting function for the stronger spectral lines (typical not valid for EW $>$ 200 m\AA). The important point is that we should be consistent in order to reduce the errors on the line-ratios.

\begin{figure}[!ht]
\centering
\includegraphics[width=8cm]{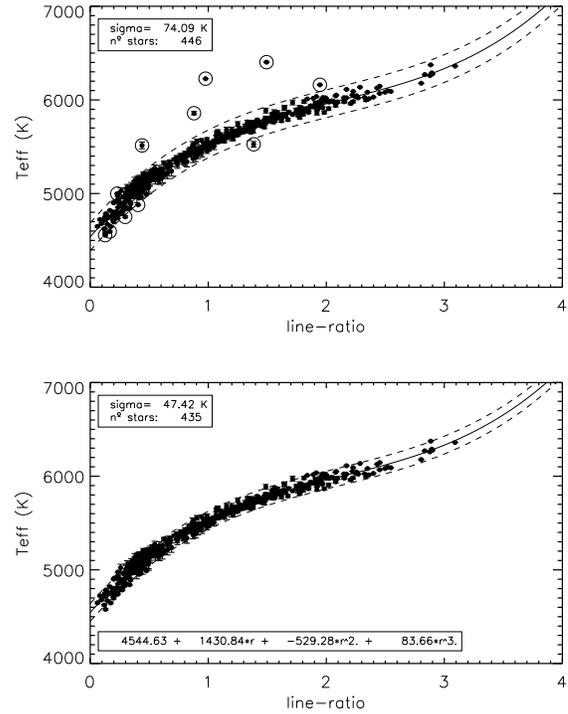}
\caption[]{Progress in the calibration of the ratio composed of the lines: SiI(6142.49\AA) and TiI(6126.22\AA). The dashed lines represent the 2 sigma from each fit. The open circles represent the points that were removal in the process. The final points and respective calibration are plotted in the bottom. }
\label{fig_calibration_proc}
\end{figure}

ARES is a fully automatic code and in the present version there is still no chance to know if the EWs were automatically well measured or if there was some error made by the code leading to an output of a bad EW measurement. Since we do not have a prior knowledge of this situation, i.e. we do not know for each lines we have a good or a bad measurement, we must first make an outlier removal. This is performed assuming that all the points in the calibration will follow a trend that can be fitted by a 3rd order polynomial function. At this point we fit all the points in the relation ''Teff vs. Ew line-ratios'' with a preliminary polynomial function than can be then used to remove the points above a specific threshold. In this case we used a value of 2 sigma for the outlier removal. This will remove the outlier points that are mainly present due to the bad measurement of one or both lines in each of the line-ratios for a given calibration star. Without these outliers we can then make a final polynomial fit that will gives us the calibration for the line-ratio. In each line-ratio we will then take into account the standard deviation and the number of stars used in the final calibration. Figure \ref{fig_calibration_proc} shows the preliminary fit to all the stars for a specific line-ratio composed of the lines SiI(6142.49\AA) and TiI(6126.22\AA). In the bottom plot we can see the final calibration fit and in the upper plot the open circles represent the the points that were removed in the process.

\begin{figure*}[ht!]
\centering
\includegraphics[width=18cm]{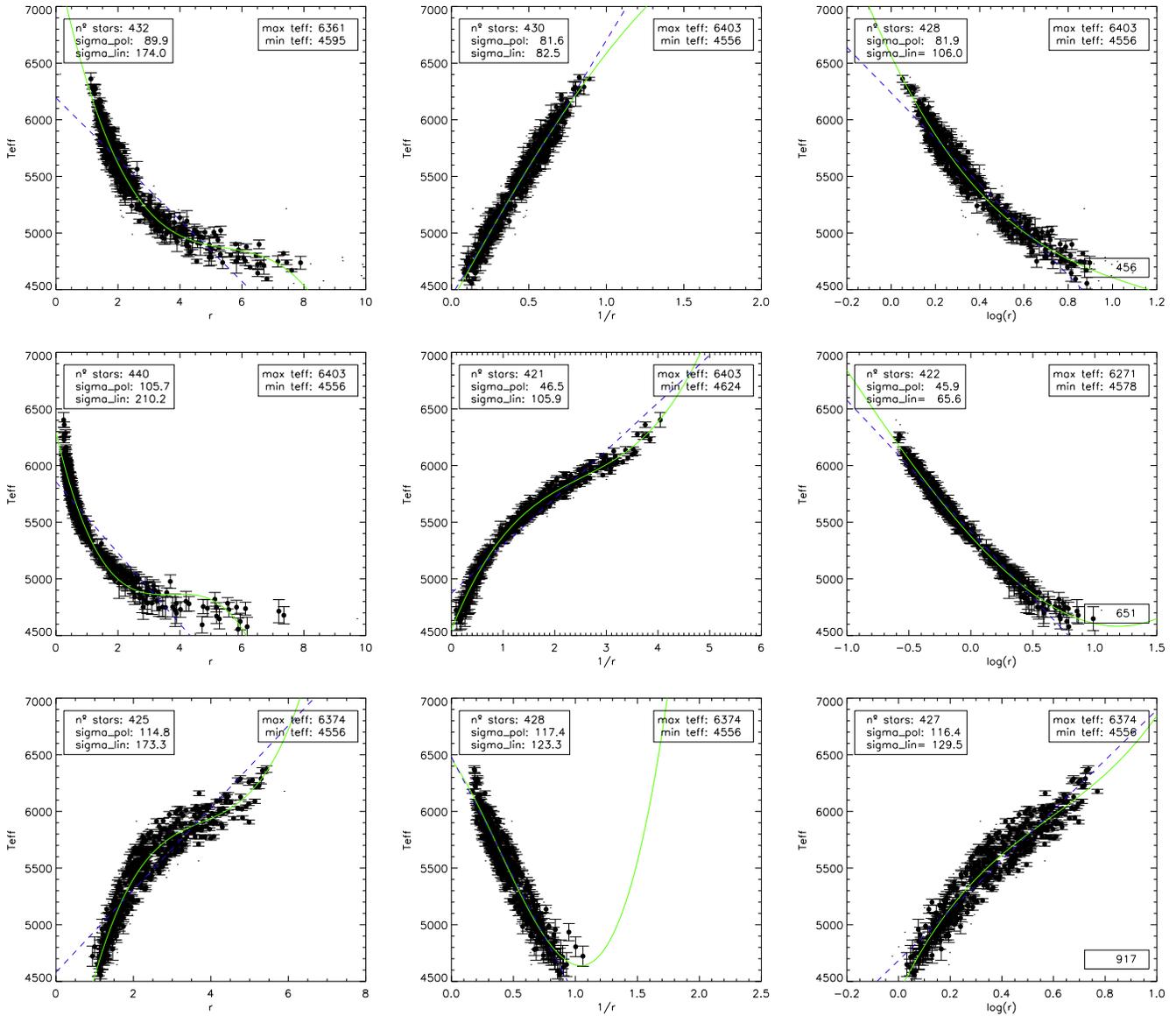}
\caption[]{Evaluation of the different fits for 3 different line-ratios (r) (from top to bottom -- line-ratios 456,651 and 917). On the left panels we plot ''Teff vs. r'', In the middle we plot ''Teff vs. 1/r''. On the right panels we plot ''Teff vs $\log(r)$''. The dashed line represents the final linear fit, and the filled line represents the 3rd order polynomial fit.}
\label{fig_calibration_proc2}
\end{figure*}

After the removal of the outliers, i.e. once we have defined which are the good line-ratio values, we make  both a linear fit  and a 3rd order polynomial function to three different plots -- Teff vs. r, Teff vs. 1/r, Teff vs. log(r) -- for a given EW line-ratio (r=$EW_i/EW_j$) and the temperature of the calibration stars (Teff). Therefore we have 6 different results for a given line-ratio which are individually derived considering the final fitting results in terms of the final number of stars used in the calibration and the standard deviation. From the six fitting results we choose the one that present the best combination of low standard deviation, large number of stars and limits in temperature. In figure \ref{fig_calibration_proc2} we show 3 line-ratio examples where we plot the final 6 different fits to the line-ratio number 456, composed of the lines CrII(4592.05\AA) and CrI(4626.18\AA) (top 3 plots), the line-ratio number 651, composed of the lines  NiI(6130.14\AA) and VI(6135.36\AA) (middle 3 plots) and the line-ratio number 917, composed of the lines FeI(6705.11\AA) and FeI(6710.32\AA) (bottom 3 plots). These 3 line-ratios represent the best, the average and the worse cases from our final list of calibrated line-ratios. In these cases different choices for the final calibration were made. In the first case we chose the straight line fit to Teff vs. 1/r, in the second case the polynomial fit to the Teff vs. 1/r, and in the last case we chose the polynomial fit to Teff vs. r. The information on the fit chosen is indicated in Table \ref{tab2} for each line-ratio so the user can know what relation should be used for each calibration.


A sample of the final line-ratio list is present in Table \ref{tab2}. This list was compiled considering the procedure described above and choosing the line-ratios that were calibrated using a number of stars greater than 300 and that present a standard deviation less than 120 K. With this criteria we have a final total number of 433 line-ratios that are built from 171 different lines.

The final line list and the final list of calibrated line-ratios is available in electronic format at http://www.astro.up.pt/... 

\begin{table*}[t]
\caption[]{Teff vs. line-ratios calibration Table. The columns correspond to the following: $num_1$ is the final ID number of the line-ration; $num_2$ is the initial ID number of the line-ration; \textit{stddev} is the final standard deviation for each assumed line-ratio calibration; \textit{nstars} is the number stars (out of a total of 451) used for the calibration of each line-ratio; $\lambda_1$ is the wavelength of the first line of the ratio; $\lambda_2$ is the wavelength of the second line of the ratio; $ele_1$ is the chemical element of the first line of the ratio; $ele_2$ is the element of the second line of the ratio; $c_0$, $c_1$, $c_2$ and $c_3$ are the fitting coefficients for each line-ratio calibration; $T_{eff}^{min}$ and $T_{eff}^{max}$ are the limits in effective temperature of the stars used for each line-ratio calibration. \textit{fit type} is the type of calibration used: the numbers 1,2,3 and 4 correspond to the relation ''Teff vs. r'' -- 3 and 4 correspond to the relation "Teff vs. 1/r" but the spectral lines are inverted in this table for ease of use --, and the numbers 5 and 6 correspond to the relations ''Teff vs. log(r)''. The numbers 2,4 and 6 correspond to 3rd order polynomial functions and the numbers 1,3,5 correspond to straight line functions. This is easily noticeable in the coefficients $c_2$ and $c_3$ (set to zero in straight line fits).}
\begin{tabular}{cc|cc|cccc|cccc|cc|c}
\hline
\hline
\noalign{\smallskip}
$num_1$ & $num_2$ & \textit{stddev} & \textit{nstars} & $\lambda_1$ & $\lambda_2$ & $ele_1$ & $ele_2$ & $ c_0$ & $c_1$ & $c_2$ & $c_3$ & $T_{eff}^{min}$ & $T_{eff}^{max}$ & \textit{fit type}\\
\hline
  ...  & ... & ... &  ...  &   ...   &    ...   &  ... &  ... &     ...    &    ...    &     ...   &     ... & \\
    92 &    189 &   92 &  406 &  6237.33 &  6274.66 &  SiI &   VI &   5052.06  &   754.45  &   169.26  &   -91.20 & 4556 & 6287 & 6 \\
    93 &    190 &   85 &  415 &  6237.33 &  6285.17 &  SiI &   VI &   5089.04  &   811.62   &  191.03  &  -113.42 & 4578 & 6361 & 6 \\
    94 &    196 &  118 &  421 &  6237.33 &  6240.65 &  SiI &  FeI &   4342.15  &  1056.95   &    0.00  &     0.00&  4578 & 6374 & 1 \\
    95 &    200 &   97 &  433 &  6243.82 &  6258.11 &  SiI &  TiI  &  4564.66  &  1335.20   &    0.00  &     0.00 & 4556 & 6403 & 1 \\
    96 &    201 &   77 &  434 &  6243.82 &  6261.10 &  SiI &  TiI  &  4486.31  &  1488.62   &   13.19  &  -142.17 & 4556 & 6403  &2 \\
    97 &    202 &   83 &  423 &  6243.82 &  6216.35 &  SiI &   VI  &  5596.97 &   1546.76  &   309.90   & -342.97 & 4556  &6289 & 6 \\
    98 &    203 &   72 &  381 &  6243.82 &  6224.51 &  SiI &   VI  &  5105.05  &   729.82   &    0.00  &     0.00  &4556 & 6136 & 5 \\
    99 &    204 &   73 &  425 &  6243.82 &  6251.83 &  SiI &   VI  &  5346.56  &   890.69  &     0.00  &     0.00 & 4556  &6276 & 5 \\
   100 &    205 &   80 &  404 &  6243.82 &  6274.66 &  SiI &   VI &   5179.51  &   809.52   &    0.00   &    0.00 & 4556&  6287 & 5 \\
   101 &    206 &   81 &  424 &  6243.82 &  6285.17 &  SiI &   VI &   5230.92  &   856.44   &    0.00   &    0.00 & 4556 & 6361 & 5 \\
   102 &    211 &  107 &  434 &  6243.82 &  6240.65 &  SiI &  FeI  &  4338.27 &   1448.16  &     0.00   &    0.00 & 4556 & 6403 & 1 \\
  ...  & ... & ... &  ...  &   ...   &    ...   &  ... &  ... &     ...    &    ...    &     ...   &     ... & \\
 
\hline
\end{tabular}
\label{tab2}
\end{table*}

\subsection{Line-ratio depths vs. Line-ratio EWs}

The typical procedure found in literature to measure the line strength in the scope of finding calibrations for the effective temperature is to use the line depth because is typically easier to measure for well chosen lines. The purpose of the work present in the paper was not to show that EWs is a better choice than the line depths. Our idea to use EWs arrives from the fact that its value should contain more information than only the depth of the line. We probably present the first work on this subject that use the EWs instead. The problems with spectral line blends and continuum determination make similar effects on the line depths and EWs determination. Since EWs have more information and since we have a powerful tool to measure these values automatically, it is clear why we made our choice.

However, ARES also has the output of the line-depth. Therefore we made a simple test for the 3 line-ratios presented in Fig 3. The result with the line-depths values shows similar or slighty higher dispersions on these 3 line-ratios (from top to bottom the dispersions of the line-ratios, in the selected plots, using line-depths are: $\sim$120 K, $\sim$53 K and $\sim$125 K). This is a very simple test that shows that the EWs are very similar to the line depth calibrations. To properly answer the question on which is the best, a detailed study on this subject must be done. This is out of the scope of this work. We think that the most important factor in these procedures, is the line list used. A proper selection of the lines should be made for each case. In our work case we used a set of lines that was proven to be stable for the measurements of EWs \citep[][]{Sousa-2008}.

\section{Using the Calibration - an automatic procedure}

\subsection{Estimating the Temperature}

In order to use the line-ratio calibration one must first measure the EWs from the stellar spectra. ARES can automatically measure the 171 lines in the list using a one dimensional spectra. Those are then used to compute the respective 433 line-ratios.

Once we have the line-ratio values we can use the calibration for each line-ratio to compute the respective individual effective temperature. Again, since we are using ARES which still does not have a control over the bad and good measurements we have to select the good line-ratios that will be then used for the final temperature estimation.

The first selection of the line-ratios is made taking out the points that gave temperatures away from the calibration interval. At this point we only accept the line-ratios that give us temperatures in the interval [4200 K, 6800 K]. This interval is chosen considering the interval in effective temperature of the calibration sample which is $\sim$ [4600 K - 6400 K] (see Fig. \ref{fig1}). We just increased this interval by 400 K in each limit in order to take into account the expected dispersion for the determination of the temperature for the stars within the limits of the interval. As an example, consider a cool star with Teff=4600K. In this case the dispersion of the line-ratio temperatures will have points below 4600 K but will be for sure above 4200K and therefore these ones will be used to get a good estimate of the effective temperature. 400K may be considered to much, but in this way we are sure that we keep the important points that would otherwise bias the temperature determination for higher values in this case.

After this first selection we can make a final outlier removal. In this procedure we choose the points that are within 2 sigma when comparing to the average temperature given by the individual line-ratios. The final estimation for the effective temperature is then obtained through a weighted average of all line-ratios that were selected using this procedure. 

\[
T_{eff} = \frac{\sum {T_{eff i}}{w_{i}} }{ \sum {w_{i}}   }
\]

The weights in each line-ratio is given as usually by ($w_i = 1/\sigma_i^2.$), where $\sigma_i$ is the standard deviation computed in each line-ratio calibration (Table \ref{tab2}). This way we are giving more strength to the best calibrated line-ratios.

\begin{figure}[bh]
\centering
\includegraphics[width=8cm]{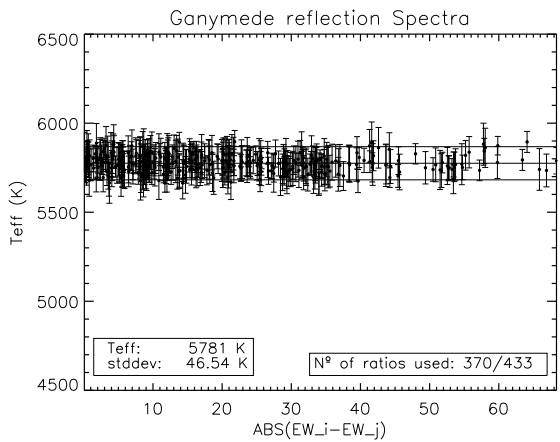}
\caption[]{Final result for the estimation of the effective temperature. Case of the solar spectrum through the reflection of Ganymede.}
\label{fig_teff_get}
\end{figure}

\subsection{Temperature Uncertainties}

The uncertainty on the final effective temperature will be obtained considering the standard deviation ($\sigma$) of the effective temperatures given by each individual line-ratio, and considering the number of line-ratios used to compute the weighted average:

\[
\Delta T_{eff} = \frac{\sigma}{\sqrt{N}}
\]

$N$ is the number of independent ratios in the final selection of line-ratios used to estimate the temperature. 

Usually in similar works, the total number of line-ratios is used instead. For multiple measurements of the same quantity, the uncertainties is derived by dividing the standard deviation by the square root of the number of independent measurements. However, all the line-ratios are not independent from each other as some have spectral lines in common. Therefore we chose the number of independent line-ratios considering the chemical elements used in each ratio. Therefore we use a smaller number $N$ that is the number of line-ratios that we consider independent. To make this clear to the reader let us consider an example where we used 3 different elements (Fe, V, Sc) and where we use 2 lines for each element. In this example we can built a total of 15 different line-ratios. For these examples we consider the number of independent line-ratios to be 3 that is given by the different combination of the elements (Fe-V, Fe-Sc, V-Sc). This is our point of view to estimate the uncertainties on the temperature, nevertheless the reader can use this calibration and choose a different approach to obtain an estimate for the errors.

The typical error for the calibration will vary from $\sim$ 10-40 K, depending on the quality of the spectra and the spectral type of the stars. Solar twins will have the lower uncertainties while the hotter and cooler stars will have the larger uncertainties, especially the cooler ones due to the increased difficulty in obtaining correct equivalent width measurements in crowder spectra range. We estimate that the systematic errors using this procedure will be below 50 K.

\subsection{Free Code to estimate Teff}

A free code was developed to use the procedure described above to estimate the effective temperature. The code is written in C++ and can be freely used in any machine. The code can be used as an extension to ARES, where ARES will automatically measures the EWs for the line list, and this code will be used to estimate the temperature. Depending on the computer speed and the type of stars, the process can derive a precise temperature in less than 1 minute, being slower for cooler stars because of the lager number of lines in the spectra. This simple code and some quick instructions are available for download on the ARES web-page: http://www.astro.up.pt/$\sim$sousasag/ares.

\begin{figure}[bh]
\centering
\includegraphics[width=8cm]{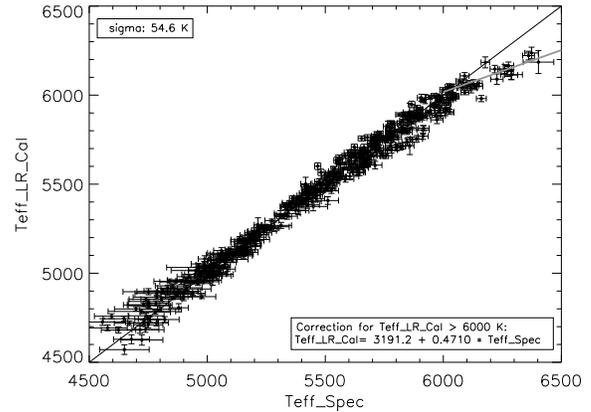}
\caption[]{Teff comparision within the sample. The offsets from the identity line at the edges are due to the limits in Teff of our sample of calibration stars.}
\label{fig_calib_teff}
\end{figure}

\begin{figure}[bh]
\centering
\includegraphics[width=8cm]{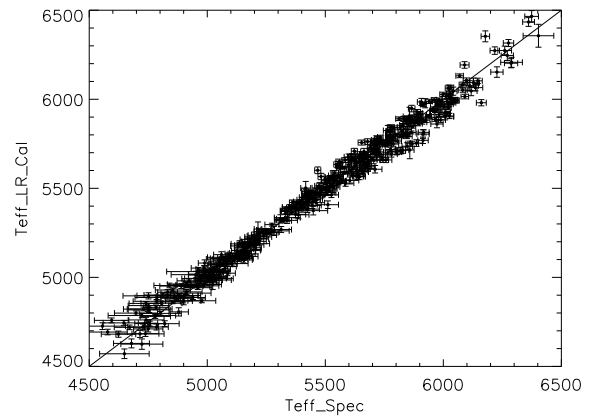}
\caption[]{Same as Fig. \ref{fig_calib_teff} but considering the small correction for hotter stars following equation \ref{eq1}.}
\label{fig_calib_teff2}
\end{figure}

\section{Testing the calibration}

\subsection{Spectroscopic Teff vs. line-ratio Teff on the sample of calibration stars}

The first test we did to the calibration was an inversion exercise where we used the sample of calibration stars to obtain the temperatures for the calibration stars. Figure \ref{fig_calib_teff} shows the comparison between the spectroscopic and the line-ratio effective temperatures. The figure shows excellent consistency except at the edges of the temperature range of the calibration stellar sample. This is due to the small number of stars in these regions when compared with the full sample (see Fig \ref{fig1}). This occurs because the fitting function for each ratio is more precise for the middle region of the relations, while at the edges it has less precision in general. Nevertheless looking at Figure \ref{fig_calib_teff} we can see that the consistence is excellent for the temperatures in the interval $\sim$ [4700-6100] K.

In order to fix the trend observed for the hotter stars we perform a small correction for the stars with temperature higher than 6000 K. The correction was achieved by fitting a straight line to these stars (see Fig. \ref{fig_calib_teff}) and use this fit to ''push'' the offset stars closer to the identity line. The following condition equations are used to make this correction for the hotter stars:

\begin{equation}
\begin{array}{lr}
T_{\rm effcor}=\frac{T_{\rm effcal}-3191.2}{0.4710} & \mbox{ if } T_{\rm effcal} > 6000 \mbox{ and } T_{\rm effcor} > 6000\\
\\
T_{\rm effcor}=T_{\rm effcal} & \mbox{ if } T_{\rm effcal} < 6000 \mbox{ or } T_{\rm effcor} < 6000

\end{array}
\label{eq1}
\end{equation}


This correction is included in the code. For the cases where the correction is applied the code outputs both the corrected and the original values of the derived temperature. Fig. \ref{fig_calib_teff2} shows a clear improvement on the results for the calibration with a small correction. We now need to test this calibration on stars outside the calibration sample. This is done in the following sections.

\begin{figure}[ht]
\centering
\includegraphics[width=8cm]{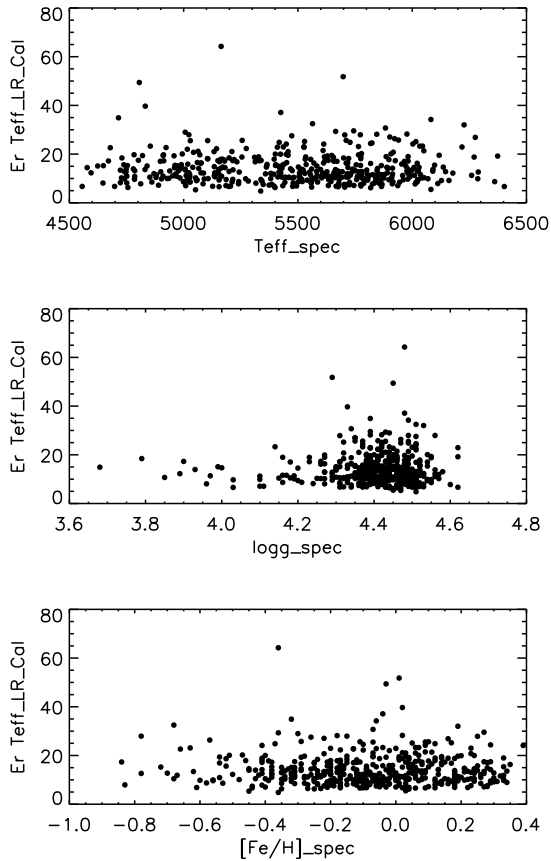}
\caption[]{Error of the temperature calibration versus stellar parameters.}
\label{fig_error}
\end{figure}

We also show the error of the temperature calibration versus the stellar parameters in Figure \ref{fig_error}. From this figure we can see that the error dependence on the stellar parameters seems to be flat indicating a clear independence with the stellar parameters. The dispersion of the error values seen in this figure should be a consequence of the error dependence on the S/N of the spectra.

\subsection{The Sun via Ganymede}
The more important test to our calibration is to use it on the Solar spectrum, estimate its temperature with the code and compare it with the canonical solar temperature of 5777 K. Figure \ref{fig_teff_get} shows this result. Using the process described previously we determine an effective temperature for the Sun using an HARPS reflection spectra of the asteroid Ganymede. The result 5782 $\pm$ 10 K  is perfectly consistent with the assumed canonical temperature of the Sun (5777 K).

\begin{figure}[bh]
\centering
\includegraphics[width=8cm]{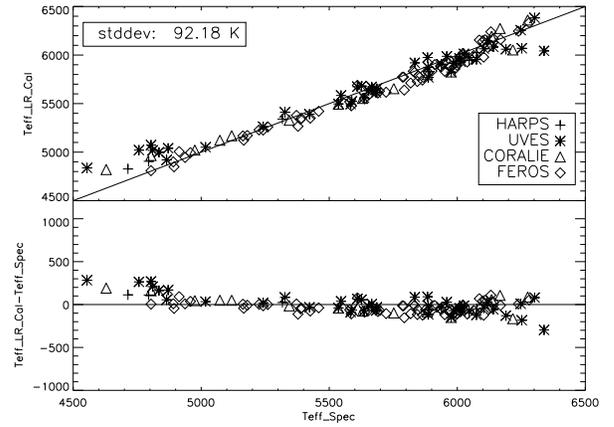}
\caption[]{Comparison of the temperature derived from the line-ratio calibration vs. the spectroscopic temperature. In this plot shows stars observed with 4 different spectrographs: HARPS, UVES, FEROS and CORALIE. Note that the Harps stars in this plot do not belong to the calibration sample.}
\label{fig_other_spec}
\end{figure}

\subsection{HARPS, UVES, FEROS and CORALIE spectra}

We used our calibration and code for different stars outside the calibration sample and observed with different spectrographs. We are using very high resolution spectra (R $\sim$ 110000) taken with HARPS and UVES and some mid-high resolution (R $\sim$ 55000) data taken by FEROS and CORALIE. This test can verify if our line-ratio calibration also works on other spectra.

This sample is composed of 15 CORALIE spectra, 7 HARPS spectra, 44 UVES spectra, and 64 FEROS spectra. The spectroscopic effective temperature for these stars were obtained in \citet[][]{Santos-2004b, Santos-2005, Sousa-2006}. The equivalent widths for this spectra were obtained using ARES. For the CORALIE and HARPS spectra the ''rejt'' parameter was varied according to the S/N of the spectra. For the UVES and FEROS spectra we took a faster approach and used the same value of ''rejt'', 0.998 for the UVES spectrograph that have typical high S/N, and 0.990 for the FEROS spectra that have a typical lower value of S/N for this specific sample. Figure \ref{fig_other_spec} shows the comparison between the spectroscopic temperature and the estimate using our  line-ratio calibration. The result is very consistent (sigma $\sim$ 90K) and shows that the calibration is valid for different spectrographs configurations and S/N. 

Since in this case we fixed the value of ''rejt'' for FEROS and UVES spectra, we also performed a test on the systematics of our calibration when using EWs computed with different values of \textit{''rejt''}. This is the same as studying the effects of systematically larger/smaller EWs on the temperatures derived from the line-ratio calibration.

\begin{figure}[th]
\centering
\includegraphics[width=8cm]{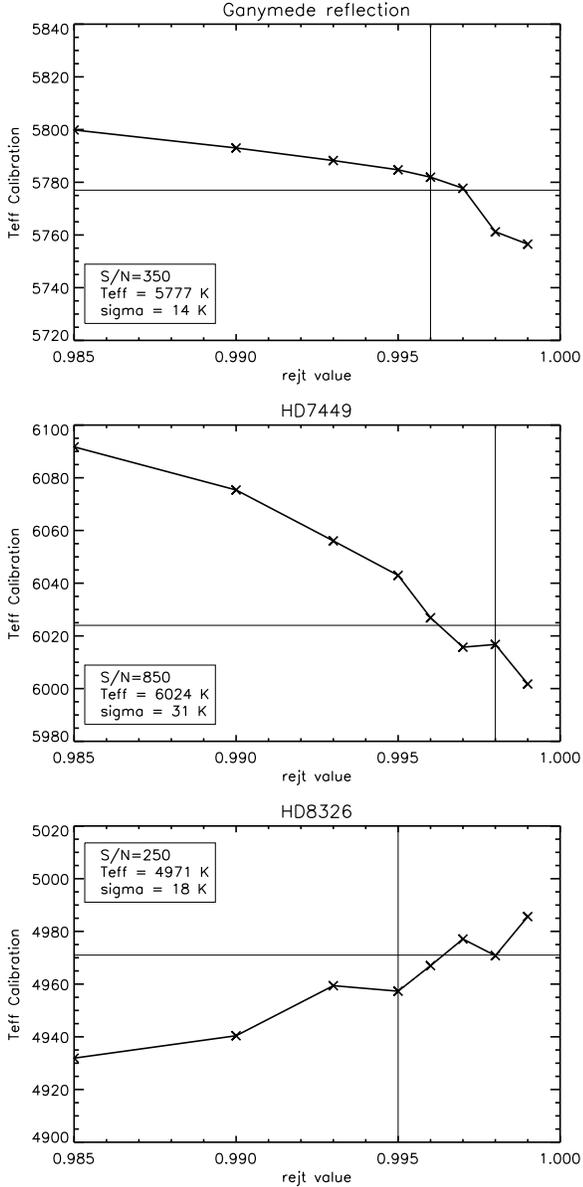}
\caption[]{Effective temperatures determination using the calibration presented in this paper when using different values of the ARES \textit{''rejt''} parameter. Three different spectral types of stars with different S/N are shown. Top: Sun through Ganymede reflection; middle: HD8326; bottom: HD7449. See text for details.}
\label{fig_''rejt''}
\end{figure}

\begin{table}[!b]
\centering 
\caption[]{Results of the calibration using different wavelength intervals. In the table we show the result of the calibrated temperature for 3 different test stars and 3 wavelength intervals. In each case $N_r$ is the number of ratios used and $N_i$ is the considered number of independent ratios.}
\begin{tabular}{lccc}
\hline
\hline
\noalign{\smallskip}
Ganymede & $T_{eff}$ & $N_r$ & $N_i$ \\
\hline
4500\AA-7000\AA: & 5782 $\pm$ 10 & 370 & 23 \\
4500\AA-5500\AA: & 5780 $\pm$ 17 &  46 &  9 \\
5000\AA-6000\AA: & 5772 $\pm$ 10 & 100 & 13 \\
6000\AA-7000\AA: & 5783 $\pm$ 12 & 241 & 17 \\
\hline
\hline
HD7449 & $T_{eff}$ & $N_r$  &$N_i$\\
\hline
4500\AA-7000\AA: & 6017 $\pm$ 11 & 335 & 22 \\
4500\AA-5500\AA: & 6015 $\pm$ 15 &  44 &  8 \\
5000\AA-6000\AA: & 6020 $\pm$ 13 & 103 & 12 \\
6000\AA-7000\AA: & 6014 $\pm$ 13 & 202 & 17 \\
\hline
\hline
HD8326 & $T_{eff}$ & $N_r$  &$N_i$ \\
\hline
4500\AA-7000\AA: & 4992 $\pm$ 12 & 380 & 23 \\
4500\AA-5500\AA: & 4993 $\pm$ 18 &  45 &  9 \\
5000\AA-6000\AA: & 5004 $\pm$ 20 & 102 & 12 \\
6000\AA-7000\AA: & 4986 $\pm$ 13 & 248 & 17 \\
\hline
\end{tabular}
\label{tab3}
\end{table}

\subsection{Propagation of the uncertainties from ARES EWs measurements}

To test how the calibration will react if we make a mistake in the continuum position for the EW measurements we used different values for the ARES parameter ''rejt''. This will put the continuum position systematically higher or lower depending on the S/N of the spectra. Fig \ref{fig_''rejt''} show the variation of the effective temperature for different values of ''rejt'', and for 3 different types of stars, the Sun, a cooler and a hotter star. The variation in temperature is expected because when we modify the continuum position (changing the ''rejt'' value) we are affecting the EW measurements. Since the weaker lines will be more affected than the stronger lines, we will have an offset in the value of the line-ratios producing a systematic shift on the derived effective temperature for all line-ratios. In Fig \ref{fig_''rejt''} we can see that the variation in effective temperature does not change by more than 40-50 K in the worse thinkable cases. This result is very significant and it means that even using different subjective measurements of the EWs, the final result for the temperature for a given star will not change by more than the 40-50 K, i.e. within the expected errors.

\subsection{Smaller wavelength intervals}

The final test we present here, is to see how the calibration will react to the use of lines in different wavelength intervals. This can be useful if you are interested in using this calibration on data that have a specific and/or a smaller wavelength coverage. To test this situation on different effective temperatures, we used the same 3 spectra as in Section 5.4: the Sun, a cooler star and a hotter star. For these 3 examples we defined 3 smaller wavelength intervals: [4500\AA-5500\AA], [5000\AA-6000\AA] and [6000\AA-7000\AA] and we tested the calibration for the lines in these intervals. The results are presented in Table \ref{tab3} and shows that the calibration works consistently well on different wavelength intervals.


\section{Conclusions}

We present a fast and automatic procedure to derive precise effective temperatures for solar-type stars. 

We selected 433 line-ratios that were calibrated using a large sample compose of 451 FGK dwarf stars. These line-ratios can be used to derive the effective temperatures with high precision using a totally automatic and fast procedure. For this purpose we built a simple code available that can be used as an extension to ARES. The calibration was tested considering different spectrographs and the results are very consistent with the temperatures that we find using a detailed spectroscopic analysis. Moreover we tested how this calibration would react to the subjectivity of ARES EWs measurements and concluded that the variation on the estimated temperature is not greater than $\sim$ 40 K. Finally we present the use of this calibration using smaller wavelength intervals and demonstrated that the temperature estimations are consistent.

\begin{acknowledgements}
S.G.S would like to acknowledge the support from the Funda\c{c}\~ao para a Ci\^encia e Tecnologia (Portugal) in the form of a grants SFRH/BPD/47611/2008. AA would like to acknowledge useful inputs on this project from Alexandra Ecuvillon and Suzanne Aigrain. NCS would like to thank the support by the European Research Council/European Community under the FP7 through a Starting Grant, as well as the support from Funda\c{c}\~ao para a Ci\^encia e a Tecnologia (FCT), Portugal, through programme Ci\^encia\,2007. We would also like to acknowledge support from FCT in the form of grants reference PTDC/CTE-AST/098528/2008 and PTDC/CTE-AST/098604/2008, POCI/CTE-AST/56453/2004, PPCDT/CTE-AST/56453/2004 and PTDC/CTE-AST/66181/2006. This research has also been supported by the Spanish Ministry of Science and Innovation (MICINN).
\end{acknowledgements}

\bibliographystyle{/home/sousasag/posdoc/mypapers/aa-package/bibtex/aa}
\bibliography{sousa_bibliography}

\end{document}